\documentclass[12pt,preprint]{aastex}
\shorttitle{The abnormally hot chromosphere and corona}
\shortauthors{LI  et al.}

\begin{document}

\title{How are the abnormally hot chromosphere and corona heated by the solar magnetic fields?}

\author{K. J. $Li$\altaffilmark{1, 2}, J. C.  $Xu$\altaffilmark{1,2},   W. $Feng$\altaffilmark{3}, J. L. $Xie$\altaffilmark{1,4}, X. J. $Shi$\altaffilmark{1},  L. H. $Deng$\altaffilmark{5}
}
\affil{$^{1}$Yunnan Observatories, Chinese Academy of Sciences, Kunming 650011, China
\email{lkj@ynao.ac.cn}}
\affil{$^{2}$State Key Laboratory of Space Weather, Chinese Academy of Sciences, Beijing 100190, China}
\affil{$^{3}$Research Center of Analysis and Measurement, Kunming University of Science and Technology, Kunming 650093, China}
\affil{$^{4}$Yunnan Key Laboratory of the Solar physics and Space Science
\email{xiejinglan@ynao.ac.cn}}
\affil{$^{5}$School of Mathematics and Computer Science, Yunnan Minzu University, Kunming 650504, China}
\email{}

\begin{abstract}
The corona is a structure possessed by stars, including the sun. The abnormal heating of the solar corona and chromosphere is one of the greatest mysteries in modern astronomy. While state-of-the-art observations have identified some candidates of magnetic activity events that could be responsible for this abnormal heating, and theoretical studies have proposed various heating modes, a complete physical picture of how they are heated as a whole remains elusive. In this study, the characteristics of the heated corona and chromosphere are investigated, and for the first time, the question of how they are abnormally heated is explicitly answered by analyzing the long-term observations of the global chromosphere in the Ca II K line and the global corona in the coronal green line.
The findings reveal that both the quiet chromosphere and corona are in anti-phase with the solar cycle, whereas the active chromosphere and corona are in phase with it.
Different parts of the solar corona and chromosphere exhibit significantly different variation characteristics, and are found to be heated by different magnetic categories and probably in different modes.
This study posits that unraveling the heating mystery is best approached through the lens of magnetic categories, rather than magnetic activity events.
\end{abstract}

\section{Introduction}
At present, the Sun is the only star for which spatial resolution has been achieved in observations, making the solar corona a primary example and benchmark for understanding stellar coronae.
In solar astrophysics, there is a longstanding and vexing question: why does the solar atmospheric temperature rise outward from the chromosphere to the corona? It is in violation of thermodynamics and thus called the abnormal heating of the upper solar atmosphere; in particular in astronomy community, the anomalous heating of the solar corona is chosen to be one of the eight key problems in modern astronomy (Kerr 2012).
So far, extensive research has been carried out to address this issue, from observations with high spatial and temporal resolutions to theoretical investigations, and many viewpoints have been put forward to partially address the issue (De Pontieu et al. 2017; Zirker $\&$ Engvold 2017; Li et al. 2019, 2022); however, a complete physical picture has not yet been established for the anomalous heating.
Observations from modern large-aperture telescopes and space satellites have demonstrated that some small-scale magnetic activity events, such as spicules and nano-flares, contribute to the heating of the upper atmosphere via mechanisms like waves or magnetic reconnections (Cranmer 2012; De Moortel $\&$ Browning 2015; Zirker $\&$ Engvold 2017; Morgan $\&$ Hutton 2018; Li et al. 2018, 2019, 2022). Nonetheless, it is not definitively proven that the heating of the whole upper atmosphere is caused mainly by these events (Judge 2021).
Many heating modes of waves and reconnections have been theoretically proposed to increase the efficiency in heating the upper atmosphere, but neither observations nor theories have identified which mode may be the main heating mechanism in the chromosphere and the corona.
This is because observing the action and identifying a certain mode on the whole for a long time are very difficult, despite the ease of finding a certain type of active events that could heat the upper atmosphere to some extent through a certain mode (Cranmer 2012; De Pontieu et al. 2017; Samanta et al. 2019; Judge 2021).

Jin et al. (2011) and Jin $\&$ Wang (2012) analyzed magnetic elements from 3764 daily magnetograms recorded from September 1996 to February 2010 by the Michelson Doppler Imager on board the Solar and Heliospheric Observatory.
In the magnetograms, as the noise of magnetic signals amplifies with the increase of the distance from the disk center, only those magnetic elements with heliocentric angles of less than $60^{\circ}$ were counted up by them.
They demonstrated that the magnetic fields on the solar surface can be divided into five categories: the large-scale magnetic fields in active regions including sunspots (designated as Category-I hereafter), and four categories of the small-scale magnetic fields outside these active regions. The flux of the magnetic fields in Category-I accounts for $47.17\%$ of the total flux of all the counted magnetic elements, but compared with other categories, the number of magnetic elements in Category-I is extremely low and negligible. The time-latitude distribution of the magnetic fields in this category is known as the butterfly diagram.
The small-scale magnetic elements with magnetic flux ranging from $(4.27 - 38.01)\times 10^{19}$ Mx, classified as Category-II, also display a butterfly diagram distribution and their temporal variation is in phase with the sunspot cycle. Category-II comprises $28.24\%$ of the magnetic flux and $16.56\%$ of the element count. Like the magnetic elements in Category-I, ephemeral active regions (Martin 1990) exhibit these characteristics as well, and here, it is speculated that these two kinds of magnetic elements may actually be the same.
Those small-scale magnetic elements whose magnetic flux is in the range of $(2.9 - 32.0)\times 10^{18}$ Mx is categorized as Category-III, which makes up $19.76\%$ in magnetic flux but as much as $77.19\%$ of the element count. They are distributed all over the solar surface, and their temporal variation is in anti-phase with the solar cycle.
They account for almost all of the network magnetic elements and may thus be called as  network magnetic elements. Combined, Category-I, -II,  and -III constitute $95.17\%$ of the magnetic flux and $92.84\%$ of the element count.
The remaining two categories of the small-scale magnetic elements show no correlation with the solar cycle. For details on the magnetic categories, please refer to Jin et al. (2011) and Jin $\&$ Wang (2012).

A consensus in the solar community is that it is the solar magnetic fields that make the solar upper atmosphere much hotter than the photosphere (Forbes 2000; Judge 2021; Li et al. 2019, 2022). Therefore, the action of the magnetic fields on the upper atmosphere is bound to leave trace(s) of ``category" in the upper atmosphere. Specifically, if the long-term evolutionary behavior opposition to the solar cycle is found, it must be the consequence of the magnetic fields in Category-III,  whereas if the in-phase occurs, it should be attributed to the magnetic fields in Category-I and/or Category-II.

At present, classic and widely used methods proceed from causes to effects to find out the activities contributing to the heating of the upper atmosphere. The observed chromosphere and corona are indeed heated. Here, we will go from effects to causes to investigate the characteristics of the heated atmosphere and then identify which categories of magnetic fields are closely related with these characteristics.
The anomalous heating is actually a full-disk phenomenon, and thus, we will analyze full-disk observations of the solar chromosphere and corona.
They have been maintaining abnormally high temperature over a long time, and thus, we will analyze their long-time observations.
The physical state of the chromosphere (density, temperature, and so on) is significantly different from that of the corona. This means that they may be heated by different objects and perhaps even in different ways. Comparative research could reveal unexpected insights, and thus we will simultaneously investigate the abnormal heating of  the chromosphere and the corona.
By employing methodologies that differ from previous studies, a comprehensive and completely new understanding to the abnormal heating is achieved, and some findings are given as well.

\section{Data and Methods}

The datasets used in the current study include 938 synoptic maps of Ca II K line intensity from Carrington rotations 827 to 1764 (August 1915 to July 1985), which were published by Bertello et al. (2020), and daily coronal index (CI) at 72 latitudes from January 1, 1939 to December 31, 2008 (Minarovjech et al. 1998), available from the website of NOAA. They are shown in Figure 1.
In the top panel of the figure, the solar latitudes from 0$^{\circ}$ to 180$^{\circ}$ are defined with the north pole at $90^{\circ}$, the equator at 0$^{\circ}$, and the south pole at $-90^{\circ}$. However, for the convenience of illustration in the bottom panel, the solar latitudes from 0$^{\circ}$ to 180$^{\circ}$ are defined with the north pole at 0$^{\circ}$, the equator at $90^{\circ}$, and the south pole at $180^{\circ}$.
In order to compare these datasets with the magnetic elements obtained by Jin et al. (2011) and Jin $\&$ Wang (2012), only latitudes less than $60^{\circ}$ will be considered in the following study.

\begin{figure*}
\begin{center}
\centerline{\includegraphics[width=1.05 \textwidth]{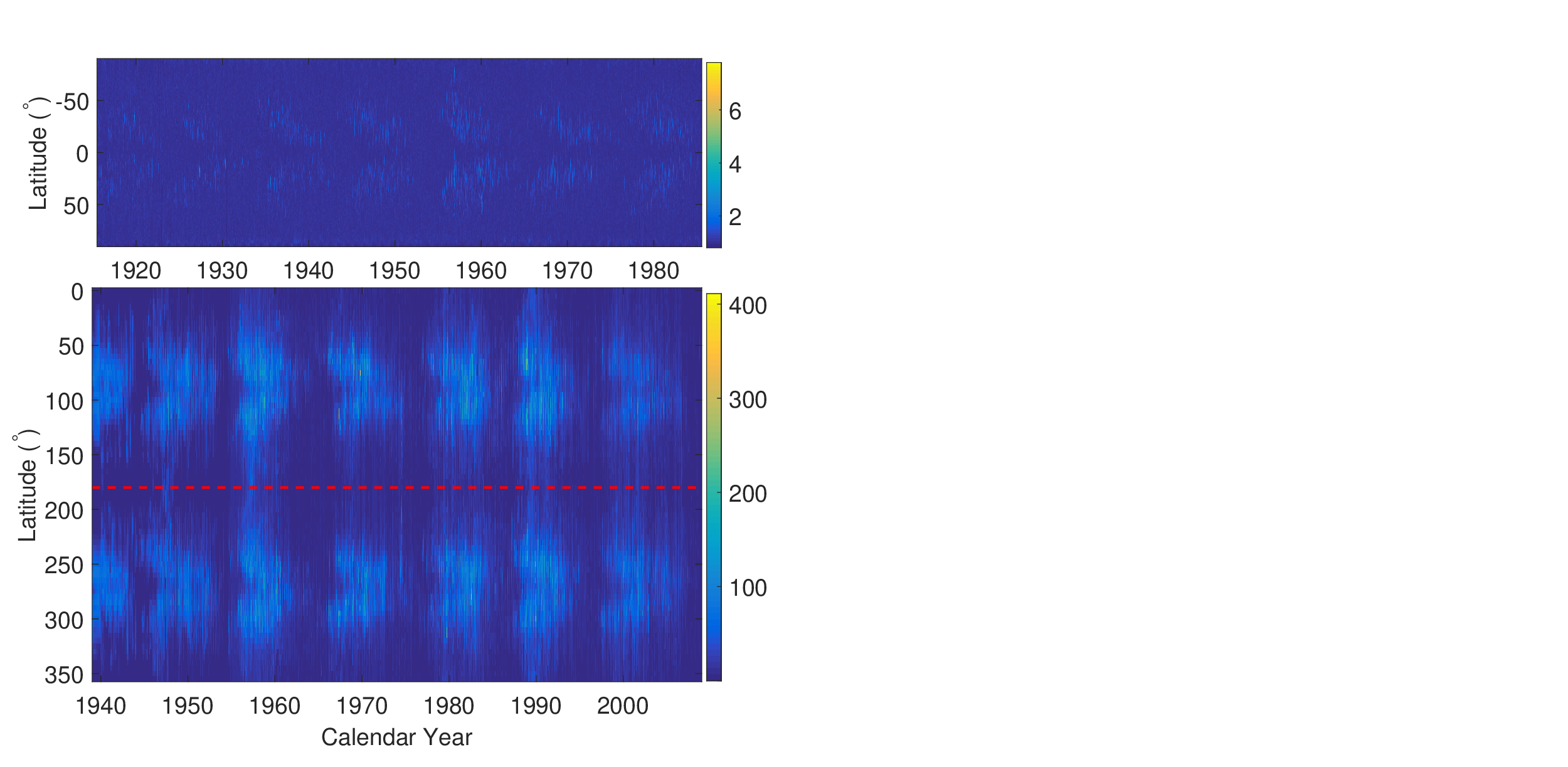}}
\caption{The top panel: synoptic maps of Ca II K line intensity from  rotations 827 to 1764 (August 1915 to July 1985).
The bottom panel: daily coronal index (CI) observed from January 1, 1939 to December 31, 2008, at the west (the area above the red dashed line) and east (the area below the red dashed line) limbs.
The latitudinal coordinates of 0$^{\circ}$--180$^{\circ}$ on the Y axis correspond to the actual solar latitudes from 0$^{\circ}$--180$^{\circ}$, ranging from the north pole ($0^{\circ}$) through the western equator ($90^{\circ}$) to the south pole ($180^{\circ}$). The latitudinal coordinates of $180^{\circ}$--$360^{\circ}$ correspond to the actual solar latitudes of $180^{\circ}$--0$^{\circ}$, extending from the south pole (Y-axis: $180^{\circ}$,  actual latitude: $180^{\circ}$) through the eastern equator (Y-axis: $270^{\circ}$, actual latitude: $90^{\circ}$) to the north pole (Y-axis: $360^{\circ}$, actual latitude: $0^{\circ}$).
 }\label{}
\end{center}
\end{figure*}

The time span of the aforementioned data ranges from August 1915 to December 2008.
The monthly mean sunspot area (SA)  is used to represent the phase of the solar cycle. SA data for this time interval are accessed through NOAA's website and are shown here in Figure 2.

\begin{figure*}
\begin{center}
\centerline{\includegraphics[width=1.05 \textwidth]{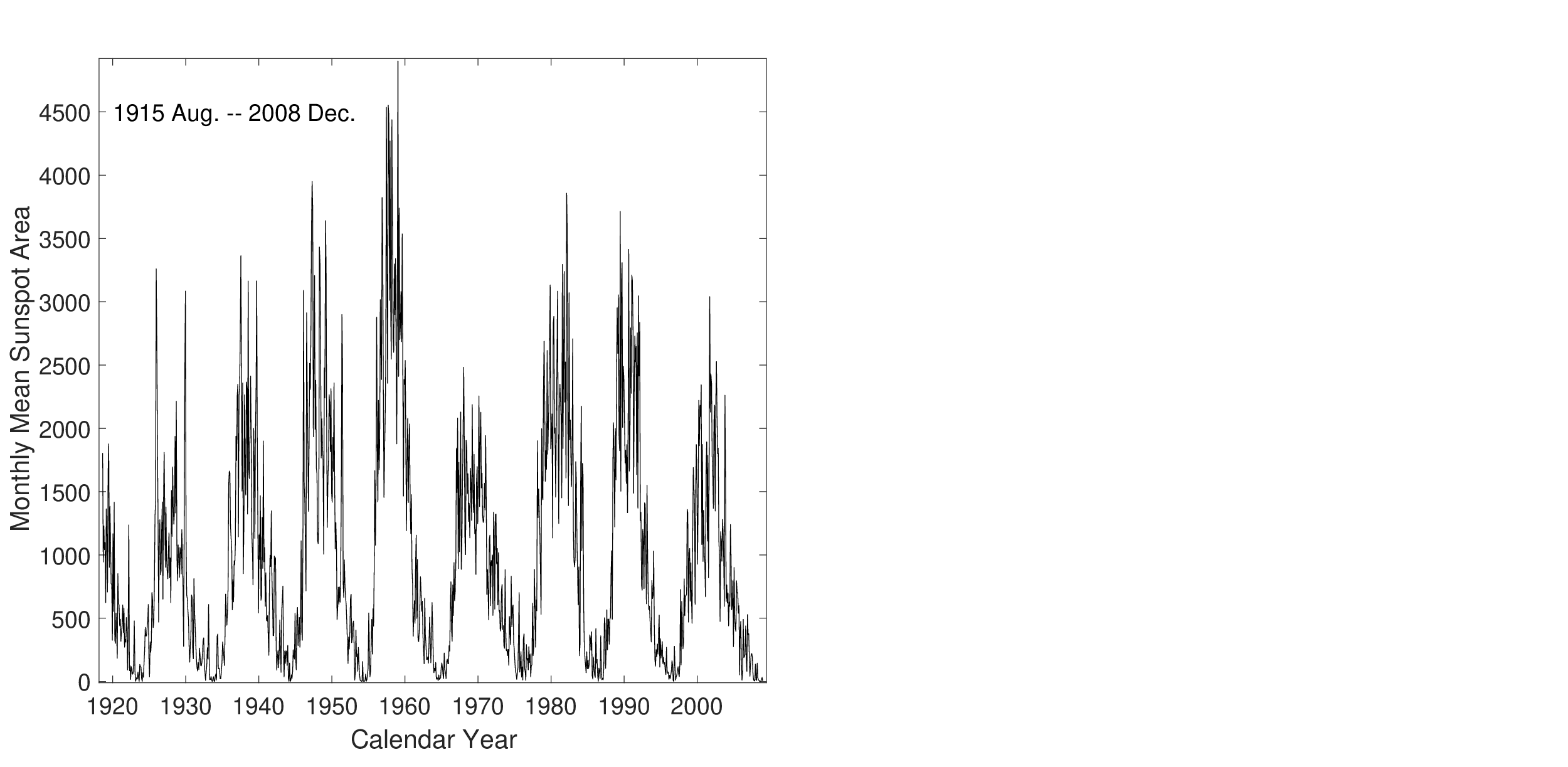}}
\caption{Monthly mean sunspot area from August 1915 to December 2008.
 }\label{}
\end{center}
\end{figure*}

The synoptic maps of Ca II intensity (Foukal et al. 2009; Bertello et al. 2014, 2020) are utilized to represent the heated chromosphere. We calculate the intensity distribution and the cumulative distribution percentage, which are displayed in Figure 3. The cumulative distribution percentage at a certain intensity is defined as the percentage of data points that are less than or equal to that intensity out of all data points.

\begin{figure*}
\begin{center}
\centerline{\includegraphics[width=1.05 \textwidth]{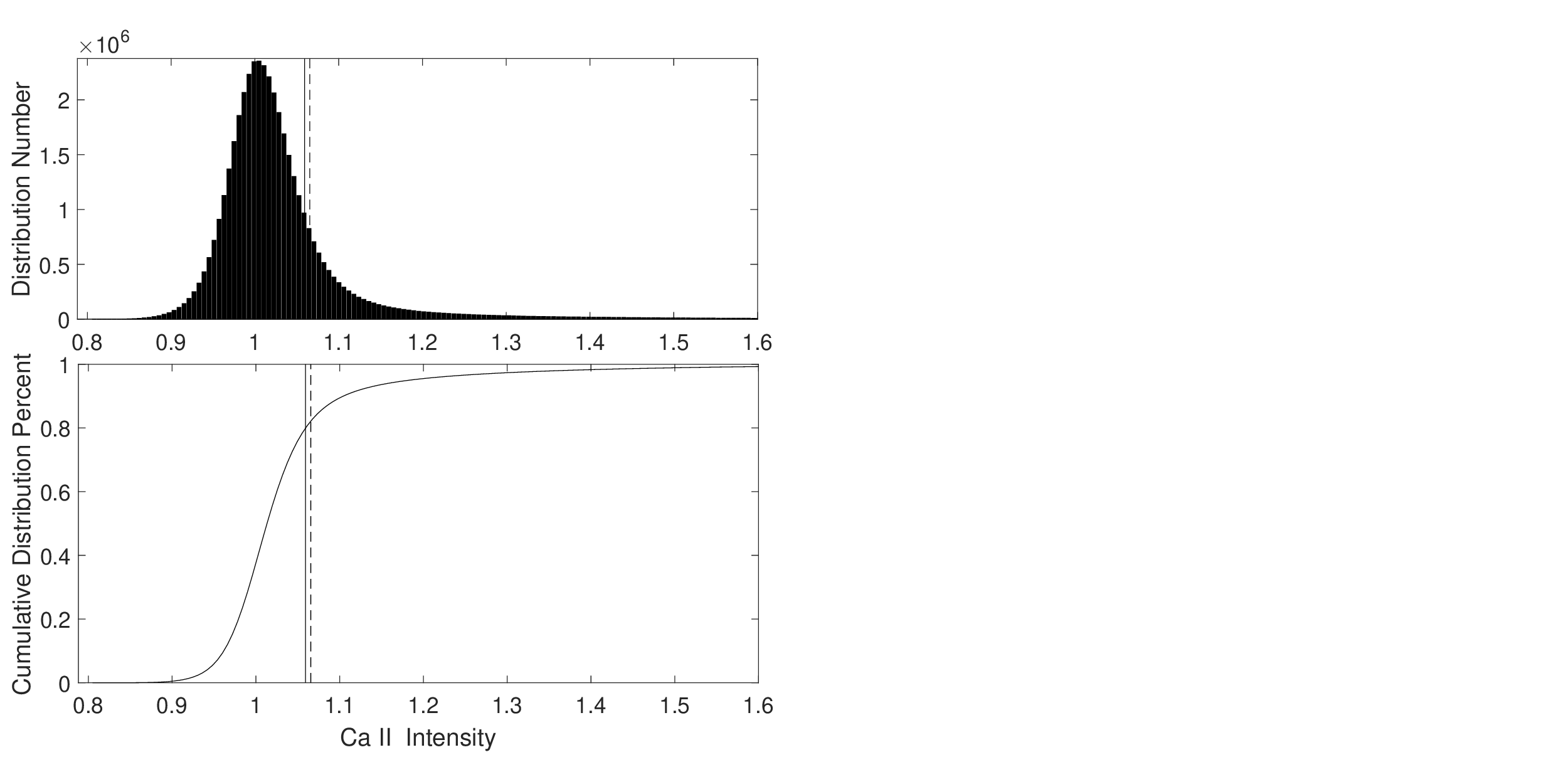}}
\caption{The distribution histogram (the top panel) and the cumulative distribution percentage (the bottom panel) of Ca II K line intensity in the 938 synoptic maps from Carrington rotations 827 to 1764. The vertical solid (dashed) line is located at an intensity of 1.0592 (1.0655).
 }\label{}
\end{center}
\end{figure*}

The corona was routinely observed in the green coronal line (5303 $\AA$) by five ground-based coronagraphs (Rybansky 1994a, 1994b), and a time series of daily coronal index (CI) at each of 72 latitudes for the period from January 1, 1939, to December 31, 2008, has been constructed at the Lomnicky Stit Station, based upon the observations of these coronagraphs  (for details, please refer to Minarovjech et al. 1998, 2007).
The CI is a homogeneous coronal data set (Rusin et al. 2004), and here it is utilized to represent the heated corona.
Although CIs are divided into east and west limbs in Figure 1, they will be considered collectively in the following study, without distinguishing between the east and west limbs.
Similarly, we calculate the distribution and the cumulative distribution percentage of CI, which are displayed in Figure 4.

\begin{figure*}
\begin{center}
\centerline{\includegraphics[width=1.05 \textwidth]{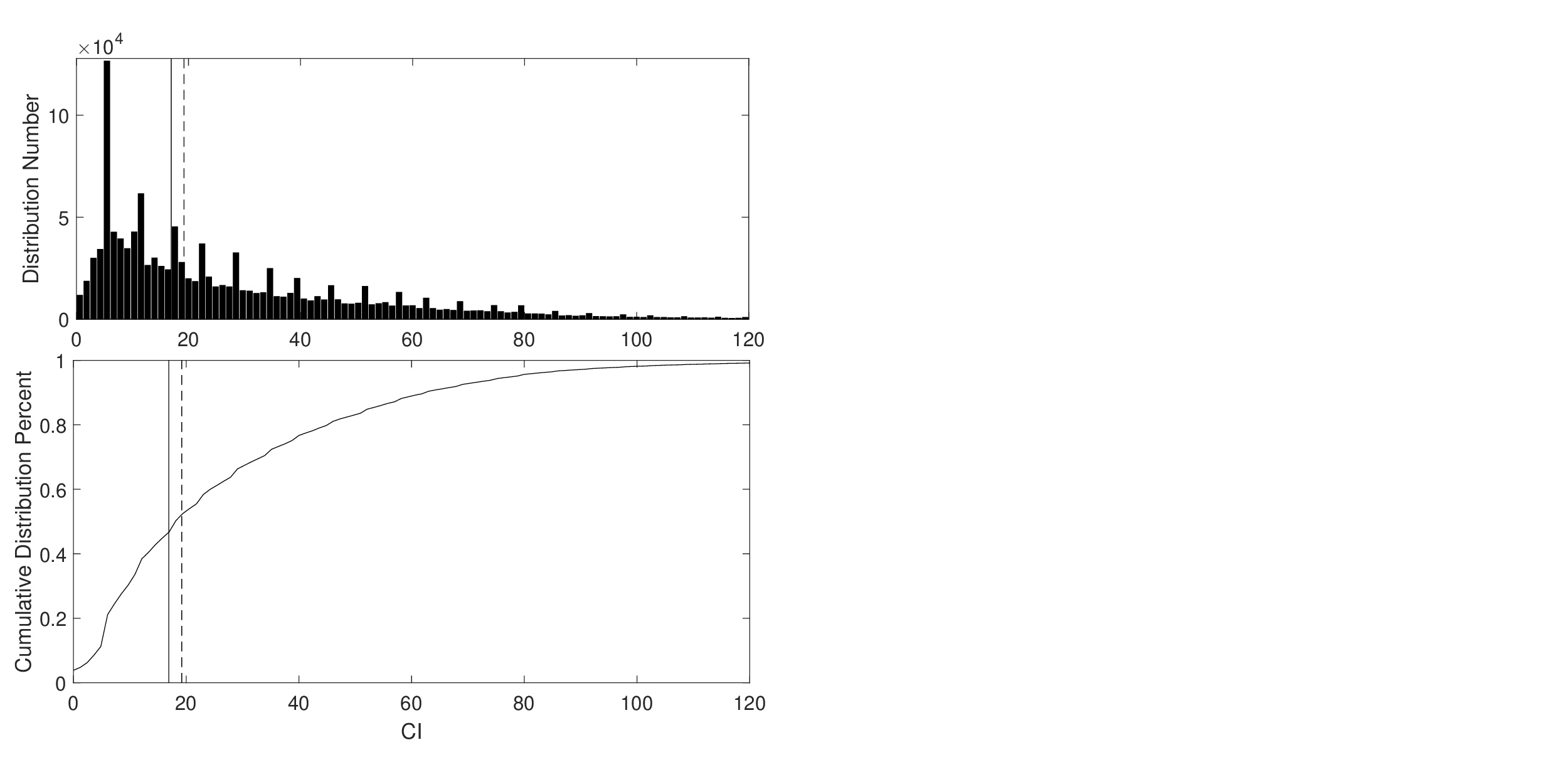}}
\caption{The distribution histogram (the top panel) and the cumulative distribution percentage (the bottom panel) of CI during the time interval of  Jan 1, 1939 to Dec 31, 2008. The vertical solid (dashed) line is located at CI of 16.9166 (19.2056).
 }\label{}
\end{center}
\end{figure*}

All values of Ca II K line intensity are divided into 200 intervals: ($x_{i}, x_{i+1}$), where i=1, 2, 3, ..., 200. Let $x_{i}=0.96+(i-1)\times 0.006$ for i=2, 3, ... 200, and $x_{1}$ is assigned the minimum value (0.8053) of these intensity data, while $x_{201}$ is assigned the maximum  value (5.9584). The Carrington rotation time is converted to calendar time.
Then, for an interval ($x_{i}, x_{i+1}$), we count the number ($N_{i}$) of data points in each month within the interval, and a time series of the monthly number of data points with intensities in the interval can be obtained.
The correlation coefficient (CC) of $N_{i}$ and SA is calculated and regarded as the CC of the average of the two endpoints of the interval, ($x_{i} +x_{i+1}$)/2. 200 values of CC can be obtained for the 200 intervals.
Figure 5 shows CC varying with ($x_{i} +x_{i+1}$)/2. Similarly, we calculate the total intensity (Total$_{i}$) of the data points in each month, whose intensity values are located in an interval ($x_{i}, x_{i+1}$). Then the CC of Total$_{i}$ and SA is calculated, representing the CC at the intensity value ($x_{i} +x_{i+1}$)/2, which is displayed in the figure as well.
Approximately 840 pairs of data participate in the CC calculation each time, and the $95\%$ confidential level line is shown in the figure.
As the figure shows, those data points  with intensity values smaller than 1.0592 are in anti-phase with the solar cycle (called Ca$_{anti}$ hereafter), while the data points with intensity values greater than 1.0655 are in phase with the solar cycle (called Ca$_{in}$ hereafter).
The correlation analysis of SA with both $N_{i}$ and Total$_{i}$ shows the same result.

\begin{figure*}
\begin{center}
\centerline{\includegraphics[width=1.05 \textwidth]{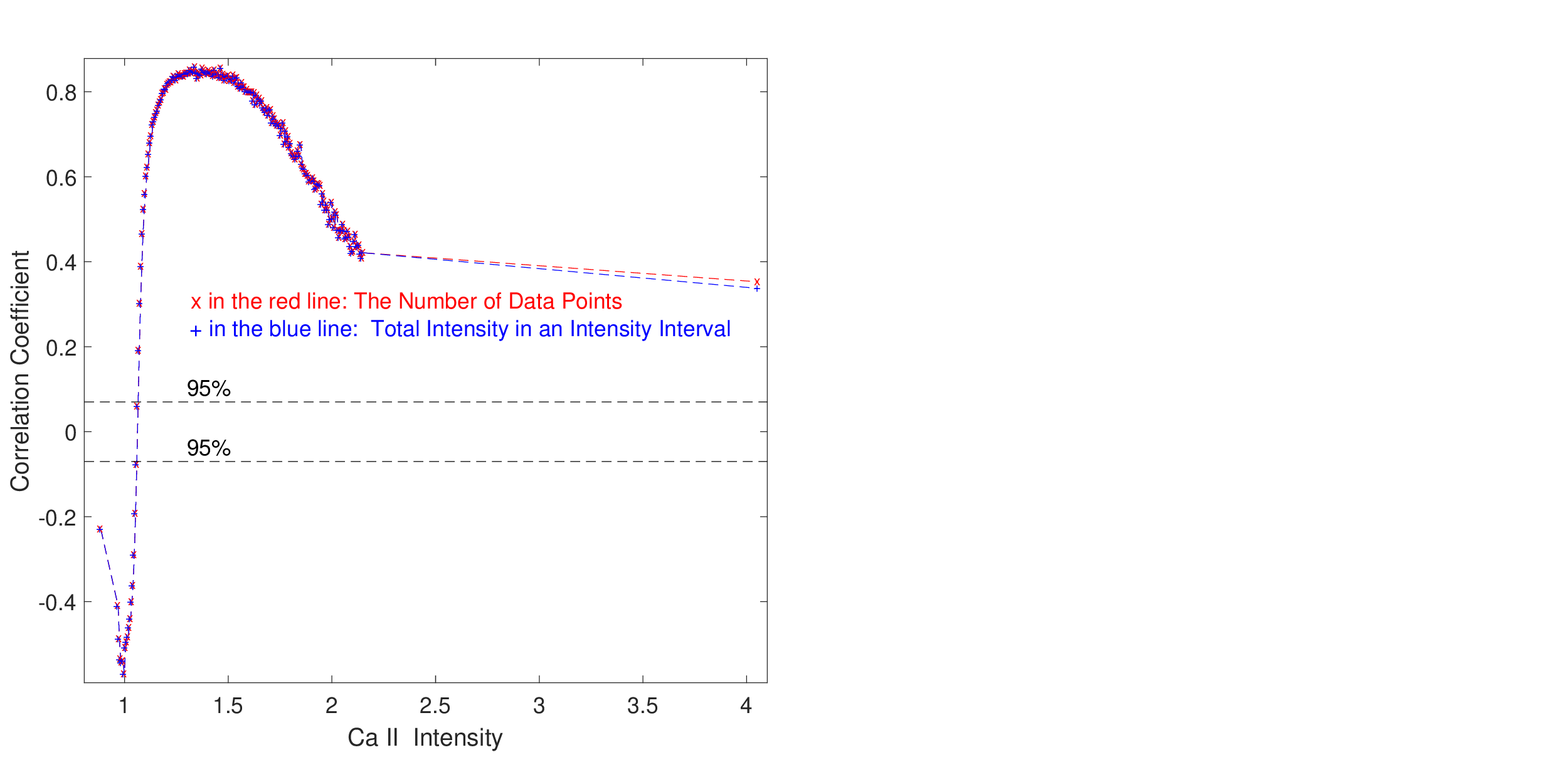}}
\caption{Correlation coefficient (red crosses in a dashed red line) of the monthly number of data points in an intensity interval with the monthly  mean sunspot area, and that (blue pluses in a dashed blue line) of the monthly total intensity value in an intensity interval with the monthly  mean sunspot area. The horizonal dashed black lines indicate the $95\%$ confidence level.
 }\label{}
\end{center}
\end{figure*}

Based on Figure 5, Ca II intensity can be divided into two special parts: the part (1.0655--5.9584) in which intensity values are  positively correlated with the solar cycle, and the part (0.8053--1.0592) in which intensity values are negatively correlated with the solar cycle.
For the former, the CC of Total$_{i}$ and SA is calculated to be 0.7902, while the CC of $N_{i}$ and SA is 0.7745. At this moment, all intensity values within this part are considered. In the same manner for the latter, the CC of Total$_{i}$ and SA is $-$0.5529, and  the CC of $N_{i}$ and SA is $-$0.5558.

Similarly, we divide all CI values into 100 intervals: ($y_{j}, y_{j+1}$), where j=1, 2, 3, ..., 100. $y_{j}=5.0+(j-1)\times 1.1$ for j=2, 3, ..., 100, and $y_{1}$ is equal to the minimum value (0) of all the considered CI data, while $y_{101}$ is the maximum value (411.0).
Then, for the monthly number ($N_{j}$) of the data points whose CI values are within an interval, ($y_{j}, y_{j+1}$) , a correlation analysis with SA is conducted, and the obtained CC is shown in Figure 6.
The total CI value (Total$_{j}$) of the data points within an interval ($y_{j}, y_{j+1}$) for a given month is also subject to a correlation analysis with SA, and the obtained CC is shown also in the figure. These two kinds of correlation analyses show the same result: CIs smaller than 16.9166 are in anti-phase with the solar cycle (called CI$_{anti}$ hereafter), while CIs larger than 19.2056 are in phase with the solar cycle (called CI$_{in}$ hereafter).

\begin{figure*}
\begin{center}
\centerline{\includegraphics[width=1.05 \textwidth]{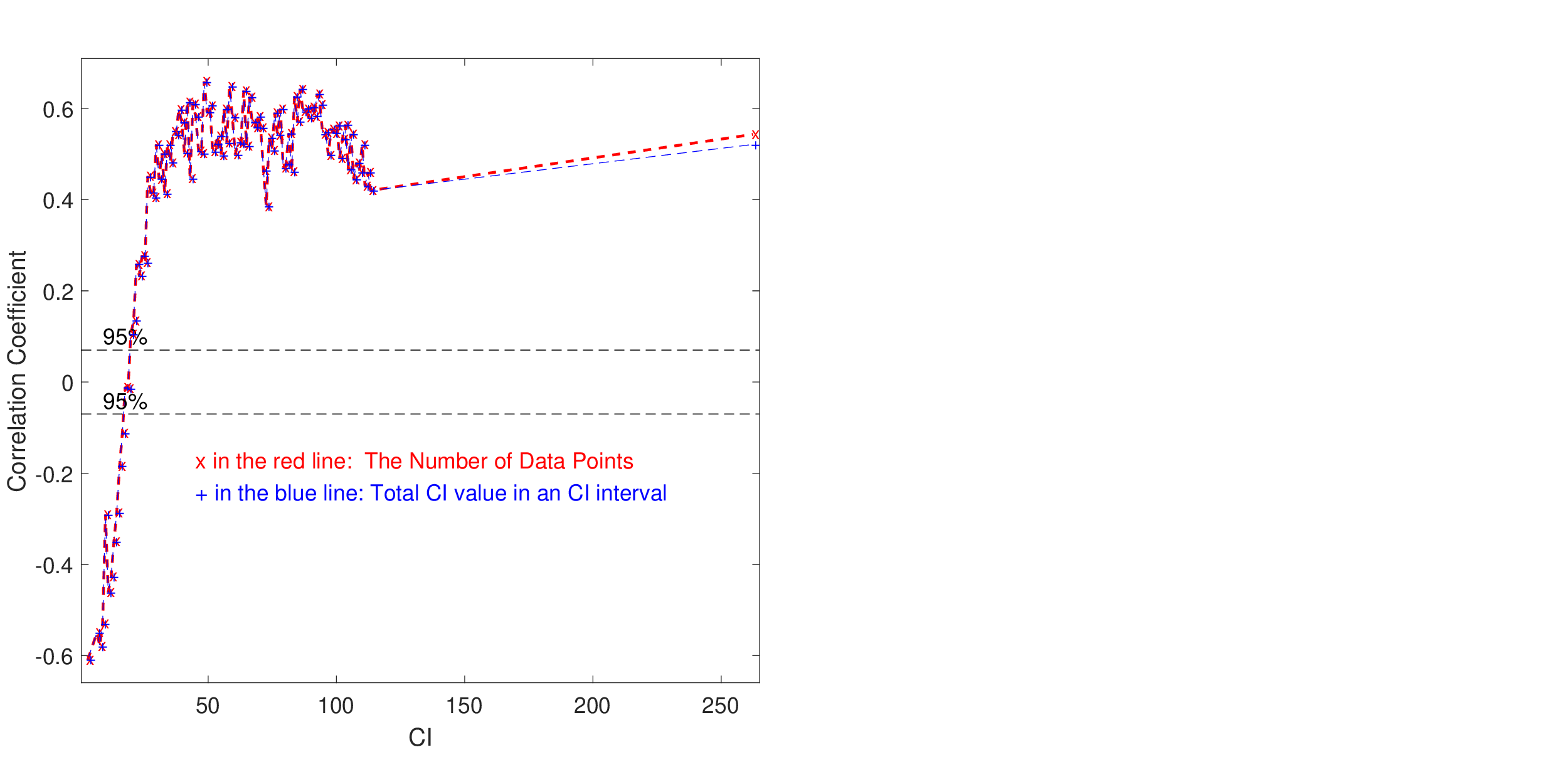}}
\caption{Correlation coefficient (red crosses in a dashed red line) of the monthly number of data points in a CI interval with the monthly  mean sunspot area, and that (blue pluses in a dashed blue line) of the monthly total CI value in a CI interval with the monthly  mean sunspot area. The horizonal dashed black lines indicate the $95\%$ confidence level.
 }\label{}
\end{center}
\end{figure*}

Based on Figure 6, CI values can be divided into two distinct parts: the part (19.2056--411) in which CI values are positively correlated with the solar cycle, and the part (0--6.9166), which is negatively correlated with the solar cycle.
For the former, the CC of Total$_{j}$ and SA is 0.8212, and the CC of $N_{j}$ and SA is 0.8080. At this point, all CI values in this part are taken into account. For the latter, the CC of Total$_{j}$ and SA is $-$0.6933, and  the CC of $N_{j}$ and SA is $-$0.7895.

\section{Results}
Magnetic reconnection is characterized as the rapid release of a large amount of energy in a short period of time, namely, a sharp variation in brightness. The general impression given by observations is that,  in the long run the time and spatial scales of reconnection heating are shorter than those of wave heating on the full solar disk, but reconnection heating is stronger (Yuan et al. 2023).

The solar chromosphere and corona can both be divided into two parts: the quiet one (Ca$_{anti}$  and CI$_{anti}$) and the active one (Ca$_{in}$ and CI$_{in}$), with the intensities of the former being significantly smaller than those of the latter, respectively, in the chromosphere and corona.
As noted above, the solar magnetic fields make the solar upper atmosphere to be much hotter than the photosphere. Therefore, the long-term evolution of the quiet chromosphere and that of the quiet corona, both found to be in anti-phase with the solar cycle,  must be the consequence of the magnetic fields in Category-III. In contrast, the long-term evolution of the active chromosphere and that of the active corona, which are in phase with the solar cycle, are certainly the consequence of the magnetic fields in Category-I and -II. Li $\&$ Feng (2022) found that the long-term evolution of  the full-disk quiet chromosphere is in anti-phase with the solar cycle through analyzing the Ca II intensity data at different latitudes, supporting these results in terms of spatial distribution.
As mentioned earlier,  the solar full-disk magnetic fields are mainly composed of Category-I, -II, and -III. As Figures 3 and 4 show, the data points of Ca$_{anti}$ and Ca$_{in}$ account for the vast majority of the corresponding total in the chromosphere, while the data points of CI$_{anti}$ and CI$_{in}$ account for the vast majority of the corresponding total in the corona as well.
Therefore, the classification of the solar magnetic fields has a good correspondence with the heating of each part of the upper atmosphere through solar-cycle phases. Hence, it is a good perspective to use the magnetic categories to understand the abnormal heating.

Based on Figure 3, in the chromosphere, the total intensity of the Ca$_{anti}$ components
is calculated to be 3.787 times the sum of the Ca$_{in}$ components. This suggests that the heating effect of Category-III is much greater than that of Category-I and -II, implying that the chromosphere should be heated mainly by the magnetic fields in Category-III.

The total area of the magnetic elements in Category-I constitutes about 5\% of the visible hemisphere over a solar cycle (see Figure 2 in Jin et al., 2011). We assume that the top 5\% components (called Ca$_{max}$ hereafter) within the considered Ca II line intensity data are solely the consequence of the magnetic elements from Category-I, in a hypothetical extreme case.
The cumulative intensity of these Ca$_{max}$ components is estimated to represent about 32.7\%  of the total intensity of the Ca$_{in}$ components. That is, even in such an extreme case, the heating effect of Category-I is still much less than that of Category-II, leading to the conclusion that the active chromosphere should be heated mainly by the magnetic elements of Category-II.

Based on Figure 4, in the corona, the total intensity of the CI$_{in}$ components is 6.871 times the sum of the CI$_{anti}$ components, implying that the heating effect reflected by the CI$_{in}$ should be much greater than that by the CI$_{anti}$, and that the corona should be heated mainly by the magnetic fields in Category-I and -II.
The mean flux of the magnetic elements in Category-II is about 7.05 times of that in  Category-III in a solar cycle, and the magnetic flux of the small-scale magnetic fields in quiet regions outside active regions is in phase with the solar cycle (Jin et al. 2011).
Therefore, the magnetic elements of Category-I and -II  mainly heat the corona.

Similarly, the most intense $5\%$ components (called CI$_{max}$ hereafter) among all considered CI data are presumed to be caused by Category-I. In such an extreme case, the total intensity of the CI$_{max}$ components accounts for merely $20.7\%$  of the total intensity of the CI$_{in}$  components. Therefore, the active corona should be heated mainly by Category-II.
Compared to other magnetic field components,  the role of the magnetic fields in Category-I is more inclined towards explosion rather than heating, as they occupy a small area on the solar surface and occur at a low frequency.

The magnetic field lines become increasingly horizontal from the photosphere to the upper atmosphere, forming canopies (Gabriel 1976; Wedemeyer-Bohm et al. 2009), and the ubiquitous horizontal fields are also found in the quiet Sun by the Solar Optical Telescope/Spectro-Polarimeter  aboard the Hinode (Lites et al. 2008).
In the upper chromosphere, almost all particles are ionized. Here the horizontal magnetic field lines
are proposed to act like a dam to prevent charged particles from crossing the magnetic lines themselves.
They effectively block the outflow of charged particles, significantly impede energy transfer through convection and conduction, and even obstruct some waves from the top of the chromosphere.
This is inferred to be the reason why atmosphere density takes an abrupt nosedive from the chromosphere to the transition region.
The variation in particle density with atmospheric height is mainly determined by gravity and atmospheric temperature; as indicated by the VAL atmosphere model  (Vernazza et al. 1981), particle density is $10^{22} \sim 10^{23}$ m$^{-3}$ in the photosphere, several orders of magnitude higher than in the chromosphere, where it is $10^{17} \sim 10^{22}$ m$^{-3}$.
The energy related to the magnetic fields, mainly including the blocked energy and the wave-heating energy of the magnetic fields as a waveguide, can raise the chromosphere temperature to much higher values than that of the photosphere.
In order to maintain the force balance between the photosphere and the chromosphere, the chromosphere has high temperature and low density, while the photosphere has high density and low temperature.
This may be an important reason why the atmospheric temperature increases relatively mildly from the top of the photosphere to the chromosphere, and this is deduced to be the main reason for the formation of the quiet chromosphere.
Relative to the umbrella-handle configuration of the magnetic field lines in the lower atmosphere, their canopy configuration in the upper atmosphere greatly increases the chance of contact with one another, suggesting that the likelihood of magnetic reconnections occurring in the upper atmosphere should be higher than that in the lower atmosphere.
Observations indicate that both wave and magnetic reconnection can be modes of coronal heating, but magnetic reconnection is  less reported in the chromosphere than in the corona (De Moortel $\&$ Browning 2015), consistent with this suggestion.

In the chromosphere, the average intensity of the Ca$_{anti}$ components is $1.003\pm0.033$, while the average intensity of the Ca$_{in}$ components is $1.186\pm 0.179$, and thus their coefficients of variations are 3.29\% and 15.09\%, respectively.
99.5\% of the Ca$_{anti}$ components are in the range of 0.880--1.059, and 99.5\% of the Ca$_{in}$ components are in the range of 1.066--2.071. Therefore, the intensity values of the Ca$_{in}$ components change little, and in particular, the intensities of the Ca$_{anti}$ components vary very little.
Such a small coefficient of variation  of the quiet chromosphere implies that,
the primary mode by which the Category-III magnetic elements directly heat the quiet chromosphere should favour waves rather than magnetic reconnections.
This is because the release of energy (increasing calcium line strength) through magnetic reconnections is known to be much  more intense than that by waves, and as mentioned above, the canopy shape of the magnetic fields suggests that the likelihood of magnetic reconnection occurring in the upper atmosphere is higher than that in the lower atmosphere. Long-term observational research is needed in the future to confirm this hypothesis.

In the corona, the mean intensity of the CI$_{in}$ components is $46.730\pm 25.158$, and that of the CI$_{anti}$ components is $7.631\pm 4.449$; therefore, the coefficient of variation of the former is $53.84\%$, and that of the latter, $58.30\%$.
Intensities of the CI$_{anti}$ components vary in the range of 0--16.9, while intensities of the CI$_{in}$ components generally vary in the range of 19.2--120; the latter vary significantly more than the former. On average, the intensity of the active corona is several times higher than that of the quiet corona.
The coefficient of variation is significantly larger in the corona than in the chromosphere, indicating that the corona is much more dynamic than the chromosphere. The intensity difference between active and quiet area is much more pronounced in the corona than in the chromosphere.
These characteristics of intensity variations and magnetic configuration in the chromosphere and corona indicate that the heating effect of magnetic reconnections is more obvious in the corona than in the chromosphere, because magnetic reconnections have a greater potential to occur in the corona than in the chromosphere, causing greater fluctuations of brightness in corona than in the chromosphere.

As mentioned above, the magnetic elements of Category-III make both the quiet chromosphere and the quiet corona to be anti-phase with the solar cycle, so it should be different modes of action by the magnetic elements that create their stratification.
It is speculated here that the magnetic elements of Category-III covering the chromosphere heat it by preventing outflow of energy,
while the quiet corona is heated by  waves and reconnections, related with these magnetic elements.
It is mainly the magnetic elements of Category-II that cause the active chromosphere and the active corona to be in-phase with the solar cycle,
but the coefficient of variation of the latter is several times larger than that of the former,
and thus there may be clearly different modes of heating action by the magnetic elements to make the active atmosphere layered. That is,
the relative importance of wave heating and reconnection heating in the active chromosphere should be  significantly different  from that in  active corona. Exploring the main heating modes of the active chromosphere and the active corona is an important topic in the future.

\section{Conclusions and Discussion}
Now, for the first time, a complete picture of the anomalous heating of all parts of the upper atmosphere can be clearly described  as follows. (1) The quiet chromosphere is heated by the magnetic elements in Category-III, mainly because their canopy structures prevent charged particles, thermal energy, and some waves from escaping from the top of the chromosphere to a large extent.
The active chromosphere is heated  by the magnetic elements in Category I and II, and by the downward propagation of the energy generated  in the corona by them. The magnetic elements in
Category-II contribute significantly more to the heating of the active chromosphere than  those in Category-I.  The heating of the quiet chromosphere constitutes the vast majority of the chromospheric heating.
(2) The quiet corona is believed to be heated mainly by the magnetic elements in Category-III.  The active corona is heated by those in Category I and II, with the latter exerting a greater effect than the former. The active corona is much more significantly heated than the quiet corona.
This is the answer to the abnormal heating of the upper atmosphere, and the heating of its different areas is well connected with the different magnetic categories.  The best way to address the issue of the abnormal heating is thus from the perspective of the magnetic categories rather than from that of solar activity events.

Now, for the first time, the general statistical view  of the heating modes of both the chromosphere and the corona can be  inferred  as follows.  The heating of the quiet chromosphere may be achieved mainly by the blocking effect of magnetic canopies, and this result needs to be verified  by observational and theoretical studies in the future.
The quiet corona is heated by waves and reconnections.
Different heating modes result in the atmosphere stratification into the quiet chromosphere and the quiet corona, although the magnetic elements in Category-III cause both to be anti-phase to the solar cycle.
The active corona is heated by waves and reconnections  (van Ballegooijen et al. 2011), and  so is the active chromosphere.
It should be different heating modes that result in the atmosphere stratification into the active chromosphere and the active corona.
The active chromosphere may also be heated by the downward propagation of the energy generated in the active corona.
Observations indicate that the heating of the corona can thermally convect downward (Madjarska et al. 2021; Bose et al. 2023), thus distinct systematic red-shifts are observed in the transition region (Peter $\&$ Judge 1999), in agreement with the conclusion.

The magnetic field configuration plays an important role in the anomalous heating, and it is the configuration and the difference in action modes of various magnetic categories that lead to the stratification of the solar atmosphere.
The main heating mechanisms of different parts of the upper solar atmosphere should differ from one another, and the
detailed heating mechanisms of all parts need to be further studied in the future.

\begin{acknowledgments}
We thank the anonymous referee very much  for careful reading of the manuscript  and helpful and constructive comments which significantly improved the original version of the manuscript.
The data of the coronal index, sunspot area,  and the full-disk synoptic maps of Ca II K line intensity
are courtesy to be
downloaded from web sites. The authors would like to express their deep thanks to the staffs of these web sites.
This work is supported by the National Natural Science Foundation of China (12373059, 12373061, 11973085, 41964007), the Basic Research Foundation of Yunnan Province (202201AS070042, 202101AT070063), the Yunnan Ten-Thousand Talents Plan (the Yunling-Scholar Project),  the ``Yunnan Revitalization Talent Support Program" Innovation Team Project, the national project for large scale scientific facilities (2019YFA0405001), the project supported by the specialized research fund for state key laboratories, and the Chinese Academy of Sciences.
\end{acknowledgments}

\end{document}